\documentclass[12pt]{article}
\def\beq {\begin{eqnarray}}
\def\eeq {\end{eqnarray}}
\def\be {\begin{equation}}
\def\ee {\end{equation}}
\def \12 {{\textstyle{1\over 2}}}

\def\Tr {{\rm Tr}}
\def\dag {\dagger}
\def\del {\partial}

\def\a {\alpha}
\def\b {\beta}
\def\e {\epsilon}
\def\d {\delta}
\def\q {\theta}
\def\D {\Delta}
\def\o {\omega}

\def\ra {\rangle}
\def\la {\langle}
\def\Tr {{\rm Tr}}

\def\bphi{\bar{\phi}}

\def \l {{\lambda}}

\begin{document}

\begin{titlepage}
\null\vspace{-62pt}

\pagestyle{empty}
\begin{center}
\rightline{CCNY-HEP-01/06}

\vspace{1.0truein} {\large \bf  Orthogonal basis for the energy
eigenfunctions}\\
\vspace{.2in} {\large\bf  of the Chern-Simons matrix model} \\
\vspace{.5in}  { Dimitra Karabali
$^{a,c}$ and  B. Sakita $^{b,c}$ \footnote{{\it e-mail addresses:}
karabali@alpha.lehman.cuny.edu, sakita@scisun.sci.ccny.cuny.edu}
}\\
\vspace{.3in}  {\it $^a$ Department of Physics and Astronomy,
Lehman College of the CUNY\\ Bronx, NY 10468}\\
\vspace {.1in}
{\it $^b$ Physics Department, City College of the CUNY\\  New
York, NY 10031}\\
\vspace {.1in} {\it $^c$ The Graduate School and
University Center, CUNY\\ New York, NY 10016}\\

\vspace{0.5in}
\end{center}
\vspace{0.5in}

\centerline{\bf Abstract}

We study the spectrum of the Chern-Simons matrix model and identify an
orthogonal set of
states. The connection to the spectrum of the  Calogero model is discussed.

\baselineskip=18pt

\end{titlepage}

\hoffset=0in
\newpage
\pagestyle{plain}
\setcounter{page}{2}
\newpage

\noindent
{\bf 1. Introduction}

Recently, Susskind \cite{susskind} proposed a description of
quantum Hall effect in terms of a noncommutative $U(1)$ Chern-Simons theory.
The fields
of this theory are infinite matrices corresponding to an infinite number
of electrons
confined in the lowest Landau level. Polychronakos, later, proposed 
\cite{poly1} a
finite matrix model as a regularized version of the noncommutative
Chern-Simons theory
in order to describe systems of finite many electrons. Although the proposed
 matrix
model seems to reproduce the basic features of the quantum Hall droplets, a
 precise
relation between the matrix model spectrum and the QHE as described by
Laughlin wavefunctions
is lacking.

A formal mapping between the states of the matrix model and Laughlin states
as
presented
in \cite{states} seems to be non-unitary \cite{poly2}, while coherent state
representations of the matrix model states produce wavefunctions with
a short distance
behavior which does not agree with the Laughlin one \cite{KS}.

On the other hand the same matrix model was introduced by
Polychronakos \cite{poly3} as
being equivalent to the Calogero model \cite{calogero}, a one-dimensional
system of
particles in an external harmonic osillator potential with mutual
inverse-square
interactions.

In this paper we analyze the spectrum of the matrix model and present a
relatively
simple way to identify an orthogonal basis of states. In doing so we make
use of
known properties of the energy eigenfunctions of the Calogero model. The
paper is organized as follows. In section 2 we briefly review the
Chern-Simons finite matrix model and its spectrum. In section 3
we analyze the eigenvalue problem and identify an orthogonal basis for
the energy eigenstates. The relation to the equivalent eigenvalue problem
of the Calogero model is discussed in section 4 and the Appendix.

\vskip .2in
\noindent
{\bf 2. Chern-Simons matrix model}

The action describing the Chern-Simons matrix model\footnote{For
clarification we would like to mention that Smolin has introduced a matrix
model, also called matrix Chern-Simons theory \cite{smolin}. Although there
are some common features, the two models are  different.}
is given by \cite{poly1}
\be
S = \int dt {B \over 2} Tr \{ \epsilon _{ab} (\dot{X} _a + i [A_0,~ X_a])
X_b + 2
\theta A_0 - \omega X_a ^2\} + \Psi^{\dag}(i \dot{\Psi} - A_0 \Psi)
\label{action}
\ee
where $X_a,~a=1,2$ are $N \times N$ matrices and $\Psi$ is a complex 
$N$-vector that
transforms in the fundamental of the gauge group $U(N)$,
\be
X_a \rightarrow U X_a U^{-1}~~,~~~~~~~~~\Psi \rightarrow U \Psi
\label{transf}
\ee
The $A_0$ equation of motion implies the constraint
\be
G \equiv -iB [X_1,~X_2] + \Psi \Psi^{\dag} -B \theta =0
\label{constraint}
\ee
The trace of this equation gives
\be
\Psi^{\dag} \Psi = N B \theta
\label{trace}
\ee

Upon quantization the matrix elements of $X_a$ and the components of 
$\Psi$
become
operators, obeying the following commutation relations
\beq
\bigl[ \Psi_i,~ \Psi_j^{\dag} \bigr] & = & \delta _{ij} \nonumber  \\
\bigl[ (X_1 ) _{ij},~ (X_2 ) _{kl} \bigr] & = & {i \over B} \delta _{il} ~
\delta _{jk}
\label{CR}
\eeq

The Hamiltonian is
\be
H = \omega ( { N^2 \over 2} + \sum A_{ij}^{\dag} A_{ji})
\label{hamiltonian}
\ee
where $A = \sqrt{B \over 2} (X_1 + i X_2)$. The system contains $N(N+1)$
oscillators coupled by the constraint (\ref{constraint}).
As explained in \cite{poly1}, upon quantization, the operator $G$ becomes  
the
generator of unitary rotations of both
$X_a$ and $\Psi$. The trace part (\ref{trace}) demands that $NB \theta $ 
being the
number operator for $\Psi$'s is quantized to an integer. The traceless 
part
of the
constraint demands the physical states to be singlets of $SU(N)$.

Since the $A_{ij}^{\dag}$ transform in the adjoint and the $\Psi_{i}^{\dag}$
transform in the fundamental representation of $SU(N)$, a purely group theoretical
argument implies that a physical state being a singlet has to contain $N l$
$\Psi^{\dag}$'s, where $l$ is an integer. This leads to the quantization of
$B \theta
=l$.

Explicit expressions for the states were written down in \cite{states}. The
ground
state being an
$SU(N)$ singlet with the lowest number of $A^{\dag}$'s is of the form
\be
\vert \Psi_{gr} \ra = \bigl[ \epsilon ^{i_1...i_N} \Psi^{\dag}_{i_1} 
(\Psi^{\dag}
A^{\dag})_{i_2}...(\Psi^{\dag} A^{\dag N-1})_{i_N} \bigr] ^l \vert 0 \ra
\label{state}
\ee
where $\vert 0 \ra$ is annihilated by $A$'s and $\Psi$'s, while the excited
states can
be written as
\be
\vert \Psi_{exc} \ra = \prod_{i=1}^{N} ( \Tr A^{\dag i})^{c_i} \bigl[
\epsilon ^{i_1...i_N}
\Psi^{\dag}_{i_1} (\Psi^{\dag} A^{\dag})_{i_2}...(\Psi^{\dag} 
A^{\dag N-1})_{i_N}
\bigr] ^l \vert 0 \ra
\label{excited}
\ee
The states (\ref{excited}) have energy $\o\left({N^2 \over 2} + l {{N(N-1)}
\over 2} +
\e\right)$, where $\e= \sum_i i c_i$. They are degenerate and the degeneracy 
is given
by the number of partitions of $\e$.

The main purpose of this paper is to identify an orthogonal basis for the 
states
(\ref{excited}).

\vskip .2in
\noindent
{\bf 3. Energy eigenfunctions, orthogonal basis}

As we shall see later, it is convenient to work in the $X$-representation.
We define the state $\vert X, \phi \ra$
such that
\be
\hat{X_1} \vert X, \phi \ra = X \vert X, \phi \ra ~~~~~~~~~~~~~~\Psi
\vert X , \phi
\ra = \phi \vert X , \phi \ra
\label{xrep}
\ee
We normalize the state such that the completeness relation is given by
\be
\int \vert X, \phi \ra e^{-\bphi\phi} d\phi d\bphi \prod_{ij} dX_{ij}
\la X, \phi \vert =1
\label{completeness}
\ee
In the $X$-representation the wavefunction corresponding to a particular 
state of
the
theory is
$\Phi (X, \bphi) = \la X, \phi \vert {\rm state} \ra$. In particular the 
wavefunction
corresponding to the ground state (\ref{state}) is of the form \cite{KS}
\be
\Phi_{gr}(X, \bphi) = \bigl[ \epsilon ^{i_1...i_N} \bphi_{i_1} (\bphi
A^{\dag})_{i_2}...(\bphi A^{\dag N-1})_{i_N} \bigr] ^l e^{- {B \over 2}
\Tr X^2}
\label{ground}
\ee
where
\be
A^{\dag}_{ij} = \sqrt{{B \over 2}} \left( X_{ij} - {1 \over B} {\del \over
{\del
X_{ji}}} \right)
\label{A}
\ee
Since (\ref{ground}) is completely antisymmetric in the $i_n$-indices, the
differential operator ${\del \over {\del X_{ji}}}$ produces a nonzero
contribution only
if it acts on the $e^{-\Tr {1 \over 2} B X^2}$ factor. We then have that
\be
\Phi_{gr}(X, \bphi) = \left( \sqrt{2B} \right)^{l N (N-1)/2} ~ \bigl[
\epsilon ^{i_1...i_N} \bphi_{i_1} (\bphi
X)_{i_2}...(\bphi X^{ N-1})_{i_N} \bigr] ^l e^{- {B \over 2} \Tr X^2}
\label{groundX}
\ee
The wavefunction corresponding to the excited state (\ref{excited}) can be
written
as a linear combination of wavefunctions of the form$
 \left(\sqrt{2B}\right)^{\sum in_i} \prod_{i=1}^{N}
( \Tr X^{ i})^{n_i} \Phi_{gr}(X, \bphi)
\label{excX}
$

Given the
constraint $G$, any physical wavefunction has to be a function of $SU(N)$
singlets made
out of the hermitian matrix $X$ and the vector $\bphi$. There are two types 
of such
invariants one can construct \footnote{For an $N \times N$ matrix $X$, the
Cayley-Hamilton theorem expresses
$X^N$ as
a linear function of $X^n$, n=1,...,N-1 with coefficients which are 
symmetric functions
of the eigenvalues of $X$.
Therefore all other invariants which involve $\bphi$'s are reduced to
$\Xi^l$ times a function of $S_n$.}
\beq
S_n (\sqrt{2B}X) & =& (\sqrt{2B})^n \Tr X^n~~,~~~~n=1,...,N \nonumber
\\
\Xi (\sqrt{2B} X, \bphi) & = &  (\sqrt{2B})^{N(N-1)/2} 
\epsilon ^{i_1...i_N}
\bphi_{i_1}
( \bphi X)_{i_2}...( \bphi X^{N-1})_{i_N} 
\label {inv}
\eeq
These can be thought of as $N+1$ independent collective variables.

Any physical wavefunction has the general form
\be
\Phi  = f(S_n) ~ \Xi ^l ~  e^{- {B \over 2} \Tr X^2}
\label{Phi}
\ee

In the $X$-representation the Hamiltonian can be written as
\be
H = {\omega \over 2} \left[ -{1 \over B} {\del^2 \over {\del X_{ij} 
\del X_{ji}}} + B
\Tr X^2 \right]
\label{HX}
\ee
We want to solve the eigenvalue problem
\be
H \Phi = E \Phi
\label{eigen}
\ee
where $H$ is given by (\ref{HX}) and $\Phi$ is as in (\ref{Phi}). Doing a 
simple
similarity transformation we get
\be
\tilde{H} f(S_n)~\Xi^l = E f(S_n)~\Xi^l
\label{eigen2}
\ee
where
\beq
\tilde{H} & = & e^{{B \over 2} \Tr X^2} H  e^{-{B \over 2}
\Tr X^2}  \nonumber \\
& = & \omega \sum_{ij} X_{ij} {\del \over {\del X_{ij}}} + {\o \over
2}N^2 - {\o \over {2B}} \sum_{ij}{\del^2 \over {\del X_{ij} \del X_{ji}}}
\eeq
The Hamiltonian $\tilde{H}$
can be written as a sum of two terms
\be
\tilde{H}= \tilde{H}_0 + \tilde{H}_{-2}
\ee
where
\beq
\tilde{H}_0 & = & \omega \sum_{ij}  X_{ij} {\del \over {\del X_{ij}}} +
{\o \over 2}N^2
\nonumber \\ \tilde{H}_{-2} & = &  - {\o \over {2B}} \sum_{ij} {\del^2 
\over
{\del X_{ij}
\del X_{ji}}}
\label{tilde}
\eeq
Since the operator $\tilde{H}_0$ essentially counts the number of $X$'s, one
 can
easily
check that
\be
\left[ \tilde{H}_0,~\tilde{H}_{-2} \right] = - 2 \tilde{H}_{-2}
\ee
In other words
\be
\tilde{H} = e^{{\tilde{H}_{-2} \over {2\o}}}~\tilde{H}_0~e^{-{\tilde{H}_{-
2} \over
{2\o}}}
\ee
This implies that if $P_k$ is an eigenfunction of $\tilde{H}_0$ then
$e^{\tilde{H}_{-2}/2\o}~P_k$ is an eigenstate of the Hamiltonian 
$\tilde{H}$.
One can
easily see that
\beq
\tilde{H}_0 ~P_k  & = & (k+ \omega {N^2 \over 2}) ~P_k \nonumber \\
\tilde{H} ~ e^{{\tilde{H}_{-2} \over {2\o}}} ~P_k  & = & (k+ \omega 
{N^2 \over 2})
e^{{\tilde{H}_{-2} \over {2\o}}} ~P_k
\label{Pk}
\eeq
Using (\ref{eigen2}), (\ref{tilde}) we see that in our case $P_k$ is of the
form
\be
P_k = J_{\{\l\}}(\sqrt{2B} X) ~ \Xi ^l (\sqrt{2B} X, \bphi)
\ee
where $J_{\{\l\}}$ is a homogeneous polynomial of the form 
(the notation will be justified later)
\be
J_{\{\l\}}(\sqrt{2B} X) = \sum_{\{n_i\}} a(\{n_i\}) \prod _i \left( \Tr(\sqrt{2B}
 X)^i
\right) ^{n_i}
\ee
such that $\sum_i i n_i + l {{N(N-1)} \over 2}  = k$.

Going back to the original eigenvalue problem, the energy eigenfunctions are
 of
the form
\be
\Phi  = e^{-{B \over 2} \Tr X^2 } e^{{{\tilde{H}_{-2}} \over {2 \o}}} 
\left( J_{\{\l\}} ~
\Xi ^l
\right)
\ee
Since
\be
e^{-{B \over 2} \Tr X^2 }~e^{{\tilde{H}_{-2} \over {2\o}}} ~ X_{ij} ~ 
e^{-{\tilde{H} _{-2} \over {2\o}}}~
e^{{B \over 2} \Tr X^2 } =
X_{ij} -
{1
\over B}{\del
\over {\del X_{ji}}}
\ee
$\Phi$ can be written as
\beq
\Phi
& =& J_{\{\l\}} \left[ \sqrt{{B \over 2}} \left( X_{ij} - {1 \over B}{\del 
\over
{\del X_{ji}}}\right) \right]   ~ \Xi ^l \left[ \sqrt{{B \over 2}}
\left( X_{ij} -
{1 \over B}{\del
\over {\del X_{ji}}}\right),~\bphi \right] ~e^{- {B \over 2} \Tr X^2} 
\nonumber \\
& = &  J_{\{\l\}} (A^{\dag}) ~ \Xi^l (A^{\dag},~\bphi) ~e^{- {B \over 2} \Tr
X^2}
\eeq
where $A^{\dag}$ is given in (\ref{A}).

There are several basis sets for the polynomials $J_{\{\l\}}$. There is a
particular one which is orthogonal. This corresponds to choosing 
$J_{\{\l\}}$'s 
to be
the Jack
polynomials \cite{jack}-\cite{macdonald}.
Although, in principle, this can be proven purely within the context of the
matrix model itself, an easier proof can be given indirectly by first 
relating the
energy eigenfunctions of the matrix model to the energy eigenfunctions of 
the
Calogero model and then using well known properties of the Calogero 
eigenfunctions
\cite{lassalle}-\cite{panigrahi}. This is shown in the following section.

\vskip .2in
\noindent
{\bf 4. Relation to Calogero model}

$X$ being a hermitian
matrix, it can be diagonalized by a unitary transformation
\be
X= U x U^{-1} ~~~~~~~~~~~~~~x_{ij}= x_i \d _{ij}
\label{diag}
\ee
The relation between the matrix model and the Calogero model is achieved by
identifying the eigenvalues
$x_i$ with the one-dimensional particle coordinates of the Calogero system.

One can show that the Laplacian of the matrix model can be written as
\be
\sum_{ij}{\del\over{\del X_{ij}}}{\del\over{\del X_{ji}}}=
{1\over\D}\sum_k{{\del^2}\over{\del x_k ^2}}\D
-\sum_{k\ne l}{{J^{kl}J^{lk}}\over{(x_k -x_l )^2}}
\label{laplacian}
\ee
where  $ \D$ is the Vandermonde determinant defined by
\be
\D \equiv det (x_i ^{N-j}) =\e^{k_1 k_2 \cdots k_N} x_{k_1}^0 x_{k_2}^1 
\cdots
x_{k_N}^{N-1}=\prod_{k<l}
(x_k -x_l )
\label{vandermonde}
\ee
and $J^{ij}$ is an operator with the following action on $U$
\be
\left[ J^{kl},~U_{ij} \right]  =   U_{ik} \d_{lj}
,\ \ \ \ \
\left[ J^{kl},~U^{-1} _{ij} \right]  = -  U^{-1} _{lj} \d_{ik}
\label{com}
\ee
For completeness we show the detailed derivation of (\ref{laplacian}) in the
appendix.

In writing down the eigenvalue equation for $H$ expressed in terms of the
eigenvalues
$x_i$ and the angular variables $U$, we
notice that the term $\sum_{k\ne l}{{J^{kl}J^{lk}}\over{(x_k
-x_l )^2}} $ acts only on the $\Xi$ dependence of the wavefunction $\Phi$ in
(\ref{Phi}). Using the particular parametrization (\ref{diag}), the
invariants
(\ref{inv}) can be written as
\beq
S_n(\sqrt{2B} X) & = & (\sqrt{2B})^n \sum^{N}_{i=1} x_i^n \nonumber \\
\Xi (\sqrt{2B}X, \bphi) & = &  (\sqrt{2B})^{N(N-1)/2} 
\epsilon ^{i_1...i_N} \bphi_{i_1}( \bphi X)_{i_2} (\bphi X^2)_{i_3}...( \bphi
X^{N-1})_{i_N} \nonumber \\
& = & (\sqrt{2B})^{N(N-1)/2} \epsilon ^{i_1...i_N} \bphi_{i_1}( \bphi UxU^
{-1})_
{i_2} (\bphi
Ux^2U^{-1})_{i_3}...(
\bphi Ux^{N-1}U^{-1})_{i_N} \nonumber \\
& = & (\sqrt{2B})^{N(N-1)/2} \det(U^{-1}) \epsilon ^{k_1...k_N} 
(\bphi U)_{k_1}(
\bphi U)_{k_2} x_{k_2}...
(\bphi U)_{k_N} x^{N-1}_{k_N} \nonumber \\
& = & (\sqrt{2B})^{N(N-1)/2} \prod_{i<j} (x_i - x_j) \prod_{i=1}^{N} 
(\bphi U)_i
\label{xi}
\eeq
This implies
\beq
J^{km} J^{mk} ~ \Xi^l & = & (\sqrt{2B})^{lN(N-1)/2} \D^l J^{km} J^{mk} 
\prod_{i=1}^
{N} \left( \bphi U \right)
^l_i
\nonumber \\
& = & (\sqrt{2B})^{lN(N-1)/2} \D^l J^{km} \prod_{i \ne k,m} \left( \bphi U
\right) ^l_i l \left( \bphi U \right)
^{l-1}_k \left( \bphi U \right) ^{l+1}_m \nonumber \\
& = & l(l+1) ~ \Xi ^l
\label{jxi}
\eeq

The Hamiltonian acting on the space of physical wavefunctions (\ref{Phi})
can therefore be
written
as
\be
H = {\omega \over 2B} \left[ -{1 \over \D} \sum_i {\del^2 \over {\del x_i^
2}} \D +
\sum_{i\ne j} {{l (l+1)} \over {(x_i - x_j)^2}} + B^2 \sum_i x_i^2 \right]
\ee
The expression $\D H \D^{-1}$ coincides with the Hamiltonian of the 
Calogero
model. This actually implies that $\D~ e^{-{B \over 2} \Tr X^2}
e^{ {{\tilde{H}_{-2}} \over 2\o}} J_{\{\l\}} (\sqrt{2B} X)~ \Xi^l 
(\sqrt{2B} X, 
\bphi)$ are energy eigenfunctions of
the Calogero model.

Using (\ref{laplacian}), (\ref{xi}), (\ref{jxi}) one can show that
\be
\D e^{-{B \over 2} \Tr X^2}
e^{ {{\tilde{H}_{-2}} \over 2\o}} J_{\{\l\}} (\sqrt{2B} X) \Xi^l
(\sqrt{2B} X, \phi)
\sim \D^{l+1} e^{-{B \over 2} \Tr X^2} ~ e^{-{\hat{O}_{L} \over 4B}} 
J_{\{\l\}} 
(\sqrt{2B} x_i)
\label{caloeigen}
\ee
where
\be
\hat{O}_L=  \sum_i {\del^2 \over {\del x_i^2}} + (l+1) \sum_{i \ne j} 
{1 \over {x_i - x_j}}\left( {\del \over {\del x_i}} -{\del \over {\del x_j}} 
\right)
\ee

Comparing (\ref{caloeigen}) to the orthogonal basis of the energy 
eigenfunctions of the Calogero model \cite{uw}-\cite{panigrahi}, we conclude 
that the polynomials $J_{\{\l\}}$ ought
to be the symmetric Jack polynomials. The inhomogeneous polynomials
$e^{-{\hat{O}_L \over 4B}} J_{\{\l\}}(\sqrt{2B} x_i)$ are the symmetric 
Hi-Jack polynomials 
\cite{sogo}
-\cite{panigrahi}
which provide an orthogonal basis for the Calogero model with the 
integration measure \cite{baker}-\cite{NUW}
\be
\D^{2l+2}~ e^{-{B\sum x_i^2}} ~\prod_i dx_i
\label{measure}
\ee

The symmetric Jack polynomials $J_{\{\l\}}$ of degree $\l$ are defined
\cite{jack}-\cite{macdonald} in terms of the monomial functions
$\prod_{i}  x_i ^{\l_i}$, where $\l = \sum_i \l_i$. Let ${\{\l\}}$
indicate the partitions $\{\l_1, ...,\l_N\}$. To a partition ${\{\l\}}$
we associate the symmetric monomial function $m_{\{\l\}} =
\sum_{P} \prod_{i}  x_i ^{\l_i}$, where the sum is over all distinct
permutations $P$. For example $m_{\{2,1\}} = \sum_{i,j} x_i^2 x_j$ for
$i \ne j$. $m_{\{\l\}} (x_1, x_2,...,x_N) = 0$ if $N$ is smaller than
the number of parts of the partition $\{\l\}$. The symmetric Jack
polynomials have the following expansion in terms of
$m_{\{\l\}}$'s:
\be
J_{\{\l\}}(x_i) = m_{\{\l\}}(x_i) + \sum_{\{\mu\} < \{\l\}} v_{\{\mu
\l\}}~ m_{\{\mu\}} (x_i)
\ee
$v_{\{\mu \l\}}$ are some coefficients which depend on $l$ and $\{\mu\} <
\{\l\}$ defines a partial ordering such that
\be
\{\mu\} < \{\l\} \leftrightarrow \mu=\l ~{\rm and}~ \sum_{i=1}^{j}
\mu_{i} < \sum_{i=1}^{j} \l_{i}~{\rm for~all~} j=1,2, \cdots ,N
\ee
Since there is no generic formula, we write down the first few symmetric
Jack polynomials, relevant for the construction of the first, second and
third excited states.
\beq
&&J_{\{0\}}=1
\nonumber \\
&&J_{\{1\}}=m_{\{1\}}
\nonumber \\
&&J_{\{2\}}=m_{\{2\}} + {{2(l+1)} \over {l+2}} m_{\{1,1\}}
\\
&&J_{\{1,1, \cdots, 1\}}= m_{\{1,1, \cdots, 1\}}
\nonumber \\
&&J_{\{2,1\}} = m_{\{2,1\}} +{{6(l+1)} \over {2l+3}} m_{\{1,1,1\}}
\nonumber \\
&&J_{\{3\}} = m_{\{3\}} + {{3(l+1)} \over {l+3}} m_{\{2,1\}} + {{6(l+1)^2}
\over {(l+2)(l+3)}} m_{\{1,1,1\}}
\nonumber
\eeq
Using these one can explicitly check that the corresponding Hi-Jack
polynomials $e^{-{\hat{O}_L \over 4B}} J_{\{\l\}}$ are orthogonal
with the integration measure (\ref{measure}).

Since $J_{\{\l\}}$'s
are symmetric, they can also be expressed in terms of 
$\prod_i (S_i)^{n_i}= \prod_i (\Tr X^i)^{n_i}$ where $\sum_i i n_i =
\l$. It  is this dependence which is implied in $J_{\{\l\}} (\sqrt{2B}
X)$ in (26) and in $J_{\{\l\}} (A^{\dag})$ in (29).

Going back to the matrix model eigenfunctions (29) and recalling that 
$\prod_{ij}
d[X_{ij}]
=\D^2[ dU ] \prod_i dx_i$, we conclude that the states
\be
\Phi_{\{\l\}} = J_{\{\l\}}(A^{\dag}) \Xi^l(A^{\dag}, \phi) 
e^{-{B \over 2}\Tr X^2}
\ee
where $J_{\{\l\}}$'s are the symmetric Jack polynomials, provide an
orthogonal basis  for the 
matrix energy eigenfunctions
\be
\int  \Phi^*_{\{\l\}} \Phi_{\{\l'\}} \ \prod_{ij} dX_{ij}d\phi d\bphi 
e^{-\phi\bphi}
= 0 ~~~~{\rm for}~~~~~~~~~{\{\l\}} \ne {\{\l'\}}
\ee
Using now eq.(10), we can write an orthogonality relation for the states of
the matrix model independent of representation, namely
\be
\la \Psi_{\{\l\}} \vert \Psi_{\{\l'\}} \ra =0  ~~~~{\rm for}~~~~~~~~~
{\{\l\}} \ne {\{\l'\}}
\label{ortho}
\ee
where $\vert \Psi_{\{\l\}} \ra = J_{\{\l\}}(A^{\dag}) \bigl[
\epsilon ^{i_1...i_N}
\Psi^{\dag}_{i_1} (\Psi^{\dag} A^{\dag})_{i_2}...(\Psi^{\dag} 
A^{\dag N-1})_{i_N}
\bigr] ^l \vert 0 \ra$

The use of the $X$-representation and the resulting connection to the 
Calogero model was very helpful in identifying the Jack polynomial 
dependence of an orthogonal basis for the excited states, but the final 
result (\ref{ortho}) is independent of representation.
\vskip .2in
{\bf Acknowledgements}

We would like to thank A.P. Polychronakos for useful discussions. We
also thank S. Meljanac and L. Jonke for pointing out a minor error in
the original version. This work was supported in part by the NSF grant
PHY-9970724 and a PSC-32 CUNY award.

\vskip .2in
\noindent
{\bf Appendix}

Using the ``polar" decomposition (\ref{diag}) for X we find
\beq
dX & = & U \left( dx + [U^{-1} dU,~x] \right) U^{-1} \nonumber \\
dX_{ij} & = & U_{ik} U^{-1}_{kj} dx_k -U_{ik}U^{-1}_{lj}(x_k -x_l) 
(U^{-1} dU)_{kl}
\label{dX}
\eeq
Using the parametrization $U = e^{i \sum_{\a} t_{\a} \theta_{\a}}$ we get
\be
(U^{-1} dU)_{kl} = \left( U^{-1} {\del U \over {\del \theta_{\a}}} 
\right)_{kl}
d\theta_{\a} \equiv e^{\a}_{kl} (\theta) d \theta_{\a}
\ee
Thus
\be
dx_k = (U^{-1} dX U)_{kk}~~~;~~~~ (U^{-1} dU)_{kl} = - {{\left(U^{-1}dX
U \right)_
{kl}}
\over {x_k - x_l}} ~~{\rm for}~~k \ne l
\ee
Using this we find
\be
{\del\over{\del X_{ij}}}= \sum _k U^{-1}_{ki}U_{jk}{\del\over
{\del x_k}}-\sum_{k\ne l}{{U^{-1}_{ki}U_{jl}}
\over
{(x_k -x_l )}} J^{kl}
\label{delX}
\ee
where
\be
J^{kl} = \sum_\a e^{kl}_\a{\del\over{\del\q_\a}}
\label{J}
\ee
and  $e^{kl}_\a$ is the inverse of $e_{kl}^\a$, such that
\beq
&& \sum_{kl}e^{kl}_{\a} (\q )e_{kl}^{\b} (\q )=\d_{\a\b} \nonumber \\
&& \sum_p e^{kl}_{\a} (\q )e_{k'l'}^{\a} (\q ) =\d_{kk'}\d_{ll'}
\eeq
Expression (\ref{J}) implies that the action of $J^{kl}$ on $U$ is as
follows
\be
\left[ J^{kl},~U_{ij} \right]  =   U_{ik} \d_{lj}
\label{com}
\ee
Further
\be
\left[ J^{kl},~J^{mn} \right] = J^{kn} \d_{lm} - J^{ml} \d_{kn}
\ee
The matrix Laplacian can now be rewritten as
\beq
\sum_{ij}{\del\over{\del X_{ij}}}{\del\over{\del X_{ji}}}
&=&\sum_{ij}\bigl[ \sum _k U^{-1}_{ki}U_{jk}{\del\over
{\del x_k}}\sum _{k'} U^{-1}_{k'j}U_{ik'}{\del\over
{\del x_k'}} \nonumber \\
&-&\sum _k U^{-1}_{ki}U_{jk}{\del\over
{\del x_k}}\sum_{k'\ne l'}{{U^{-1}_{k'j}U_{il'}}
\over
{(x_k' -x_l' )}} J^{k'l'} \nonumber \\
&-&\sum_{k\ne l}{{U^{-1}_{ki}U_{jl}}
\over
{(x_k -x_l )}} J^{kl}\sum _{k'} U^{-1}_{k'j}U_{ik'}{\del\over
{\del x_k'}} \nonumber \\
&+&\sum_{k\ne l}{{U^{-1}_{ki}U_{jl}}
\over
{(x_k -x_l )}} J^{kl}\sum_{k'\ne l'}{{U^{-1}_{k'j}U_{il'}}
\over
{(x_k' -x_l' )}} J^{k'l'} \bigr]
\eeq
The first term is $\sum_k{{\del^2}\over{\del x_k ^2}}$, the second term is
zero, the third is $\sum_{k\ne l}{1\over{(x_k -x_l )}}\big({\del\over
{\del x_k}}-{\del\over{\del
x_l}}\big)$, and the last is $-\sum_{k\ne l}{J^{kl}J^{lk}\over{(x_k -x_l
)^2}}$. Thus we obtain
\be
\sum_{ij}{\del\over{\del X_{ij}}}{\del\over{\del X_{ji}}}=
{1\over\D}\sum_k{{\del^2}\over{\del x_k ^2}}\D
-\sum_{k\ne l}{J^{kl}J^{lk}\over{(x_k -x_l )^2}}
\ee
where we used
\be
\sum_k{{\del^2}\over{\del x_k ^2}}+\sum_{k\ne l}{1\over{(x_k -x_l )}} 
\big({\del\over
{\del x_k}}-{\del\over{\del x_l}}\big)
={1\over\D}\sum_k{{\del^2}\over{\del x_k ^2}}\D
\ee
and $ \D=\prod_{k<l}
(x_k -x_l ) $.


\begin{thebibliography}{99}

\bibitem{susskind} L. Susskind, hep-th/0101029.
\bibitem{poly1} A.P. Polychronakos, {\it JHEP} {\bf 0104} (2001) 011,
hep-th/0103013.
\bibitem{states} S. Hellerman and M. Van Raamsdonk, hep-th/0103179.
\bibitem{poly2} A.P. Polychronakos, hep-th/0106011.
\bibitem{KS} D. Karabali and B. Sakita, hep-th/0106016.
\bibitem{poly3} A.P. Polychronakos, {\it Phys. Lett.} {\bf B266} (1991) 29.
\bibitem{calogero} F. Calogero, {\it J. Math. Phys.} {\bf 12} (1971) 419.
\bibitem{smolin} L. Smolin, {\it Phys. Rev.} {\bf D57} (1998) 6216; {\it Nucl. Phys.} {\bf B591} (2000) 227; hep-th/ 0006137.
\bibitem{jack} H. Jack, {\it Proc. R. Soc. Edinburg (A)} {\bf 69} (1970)
1.
\bibitem{stanley} R.P. Stanley, {\it Adv. Math.} {\bf 77} (1988) 76.
\bibitem {macdonald} I.G. Macdonald, {\it Symmetric Functions and  Hall
Polynomials}, 2nd
edition, Oxford, Clarendon press, 1995.
\bibitem{uw} H. Ujino and M. Wadati, {\it J. Phys. Soc. Japan}
{\bf 64} (1995) 2703;
{\bf 65}
(1996) 653; {\bf 65} (1996) 2423; {\bf 66} (1997) 345; T.H. Baker and P.J. Forrester
{\it Commun. Math. Phys.} {\bf 188} (1997) 175; J.F. van Diejen, {\it Commun. Math. Phys.} {\bf 188} (1997) 467.
\bibitem{lassalle} M. Lassalle, {\it C.R. Acad. Sci. Paris, t. Series I} 
{\bf 312} (1991) 725; {\bf 313} (1991) 579.
\bibitem{sogo} K. Sogo, {\it J. Phys. Soc. Japan} {\bf 65} (1996) 3097.
\bibitem{baker} T.H. Baker and P.J. Forrester, {\it Nucl. Phys.} {\bf B492}
(1997) 682.
\bibitem{NUW} A. Nishino, H. Ujino and M. Wadati, cond-mat/9803284.
\bibitem{panigrahi} N. Gurappa and P.K. Panigrahi, {\it Phys. Rev.} {\bf
B59} (1999) R2490; {\it Phys. Rev.} {\bf B62} (2000) 1943.

\end{thebibliography}
\end{document}